# Time & Citation Networks[1]

## James R. Clough and Tim S. Evans

Imperial College London, Centre for Complexity Science, South Kensington Campus, London SW7 2AZ (U.K.)

## Abstract

Citation networks emerge from a number of different social systems, such as academia (from published papers), business (through patents) and law (through legal judgements). A citation represents a transfer of information, and so studying the structure of the citation network will help us understand how knowledge is passed on.

What distinguishes citation networks from other networks is time; documents can only cite older documents. We propose that existing network measures do not take account of the strong constraint imposed by time.

We will illustrate our approach with two types of causally aware analysis. We apply our methods to the citation networks formed by academic papers on the arXiv, to US patents and to US Supreme Court judgements. We show that our tools can reveal that citation networks which appear to have very similar structure by standard network measures turn out to have significantly different properties. We interpret our results as indicating that many papers in a bibliography were not directly relevant to the work and that we can provide a simple indicator of the important citations. We suggest our methods may highlight papers which are of more interest for interdisciplinary research. We also quantify differences in the diversity of research directions of different fields.

## Background

Bibliometrics has a long tradition of dealing with citation networks from a network point of view as Price's model (Price 1965) shows. The recent explosion of interest in network analysis in other fields has led to development of existing methods and introduced many new techniques. However most network methods assume static graphs where time plays no explicit role even if the underlying data is almost always evolving. Time can be incorporated into a network representation in two main ways. If we assign a single time to each edge we have a *Temporal Edge Network*. Such networks have received considerable attention (Holme & Saramäki, 2012). For instance they form a useful representation for the pattern of communications between individuals. Alternatively in *Temporal Vertex Networks* each node carries a single time. The citation network provides a natural example of the latter as each paper has its publication date. Here then we will focus on the analysis of this second type of temporal network, using the bibliometric context of citation networks to motivate our work.

The causal structure of citations plays a central role in bibliometric analyses. At the simplest level understanding the different time scales for citation patterns seen in different research fields is known to be essential. In Price's model (Price, 1965) vertices appear in a fixed order, reflecting the order of publication of real citation networks. Price's model captures the essential nature of a citation; they are always from newer to older papers. Applying Price's growing network model to other contexts where time plays a different role makes no sense e.g. links between web pages are not constrained by the age of a web site.

The constraints imposed by time are very different from the spatial constraints. Network science has few tools specifically developed to work with temporal vertex networks. However as part of our work we adapt results found in other areas: discrete mathematics, quantum gravity, and in computer science. Bibliometrics asks very different questions about such networks so applying these ideas is not always straightforward.

Our hypothesis is that existing network measures do not account for the constraint of time. So we have embarked on a programme to develop new temporally aware network measures and to prove their utility in the context of citation networks.

**Methods and Data**

Our networks are defined such that each node has a unique time. Edges can only exist from a younger to an older node, see Figure 1. Citations between academic papers are a good example, patents and court rulings have similar citation structures. All edges are directed, but the arrow of time also ensures that such networks will have no loops (acyclic) provided you follow the direction of the edges. The formal name for such a network is a *Directed Acyclic Graph* or *DAG* for short.

In practice, citation data is not exactly a DAG but we found that citations in the 'wrong' direction form less than 1% of our data so they should have a limited effect on any conclusions. We construct a true DAG by dropping any such acausal citations.

We have used a variety of data sets in our work (Clough et al. 2015, Clough and Evans 2014). We have used citation information on the arXiv repository taken from two independent different sources. This allows us to check that our results are robust against any differences in citation extraction. First we use the KDD cup data (2003) which covers the first ten years of the hep-ph and hep-th sections (theoretical and phenomenological particle physics respectively). We have also looked at a separate version which covers all sections of arXiv up to 2013 which was derived from paperscape.org they also form a citation network.

We have also studied the citation network of around 4,000,000 US patents between 1975 and 1999 (Hall, Jaffe, & Trajtenberg, 2001). Finally we worked with the network defined by about 25,000 judgements of the US Supreme court 1754 to 2002 (Fowler & Jeon 2008).

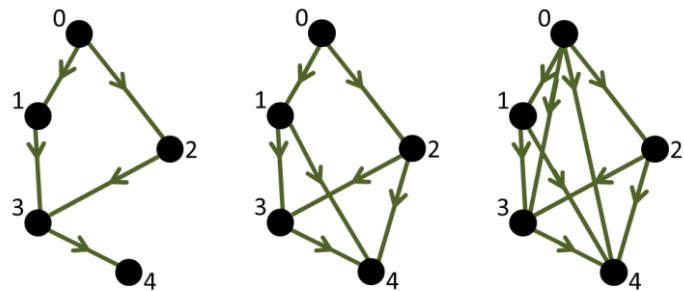

**Figure 1 The unique transitively reduction (left) and transitive completion (right) of the citation network (a Directed Acylic Graph or DAG) shown in the centre. All casual relationships implied by an edge in the central network appear as an explicit edge in the right hand network. The edges in the left hand network are the least required to capture all these causal relationships.**

*Transitive Reduction (TR)*

Our first example of a network operation which takes account of the constraint of time is Transitive Reduction (*TR*). In TR, links are removed provided that they leave the connectivity of every pair of nodes unchanged. That is if there was a path between a given pair of nodes (respecting the direction of the links) before TR, there will still be at least one such path after TR. This process can be defined on any network but for DAGs it is guaranteed to produce a unique result, see Figure 1. Algorithms for this procedure are well known in computer science but we found basic implementations in python were sufficient even for our largest networks (Clough at al. 2015)

Once we have this essential causal core of our citation network we illustrate our approach with two simple measures: the fraction of edges lost in the TR process and a comparison of the citation count of papers before and after TR.

*Dimension*

In bibliometrics, we often place papers in different fields as there is great interest in understanding the relationships between topics, as illustrated by maps-of-science (such as

Börner et al. 2012). It is natural to ask if we can assign a sense of dimension to such 'topic' spaces. A high dimension would indicate that researchers can develop work in several independent directions, a low dimension indicates that all the work in that field is tightly linked with little independence. There are some standard ways to assign an effective dimension to a network but these all assume that all directions are similar, just as moving left/right or forwards/backwards is the same for a ball on a flat table. Unfortunately, none of the measures used in the network science literature take account of time which is a very different sort of dimension. Given that temporal information is an essential part of the definition of a citation network, we must work with a different type of measure. Our work (Clough and Evans 2014).draws on inspiration from work in discrete mathematics on *posets* (partially ordered sets, e.g. Bollobás & Brightwell 1991) and from the Causal Set programme of quantum gravity (e.g. Reid 2003).

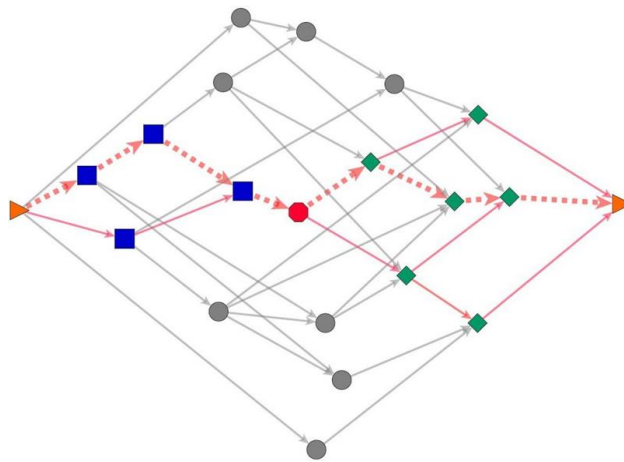

**Figure 2 An illustration of the box counting method to find dimension. Here the source and the target papers (triangles at left and right respectively) define an interval of N=19 papers - the other vertices shown here. The edges represent the transitively reduced citation network of all twenty paper. The midpoint is shown as the red circle in the centre. It defines two sub-intervals $N_1$=4 (blue squares) on the left and $N_1$=6 on the right (green diamonds). This gives D=2.16 and D=1.61 as our dimension estimates. The example was generated by throwing points down with one space and one time coordinate chosen at random, i.e. D=2.**

Our first approach is a simple box counting method (Reid 2003). We first choose a pair of papers, the source and target nodes, at random. We then find the *interval* defined by the source and target nodes, which is the set of all $N$ papers which lie on a path between source and target. As always our paths must respect the direction of time. Next we find the midpoint, a node chosen such that two sub-intervals defined by source and midpoint, and by midpoint and target nodes, are roughly equal size $N_1 \approx N_2$. It then follows that we should expect the 'length' scale of our two smaller intervals interval to be roughly half that of the large interval. Assuming papers are scattered at equal density in our data, we can use the number of points in an interval as a measure of the volume in the space-time. It then follows that the ratio of the number of points from small to large interval should scale as $N_1/N \approx N_2/N \approx 2^{-D}$. By analysing many intervals within one academic field the space-time dimension $D$ (one time and $(D-1)$ topic space dimensions) of that field may be found.

The second method we use here is the Myrheim-Meyer dimension estimator (see Reid 2003 for references). To do this we again pick a source and target paper. We then count the number causally connected pairs $P$ in the interval defined by our source and sink which contains $N$ nodes and these are related by $(P/N^2) = \Gamma(D+1) \, \Gamma(D/2) \, / \, (4 \, \Gamma(3D/2) \, )$ where $\Gamma(x)$ is the standard Gamma function. This formula is derived for a large $N$ by assuming points are sprinkled at uniform density in Minkowskii space-time. We have also used the same approach to show that in a different type of space, the cube box space of Bollobás & Brightwell (1991) the formula is simply $P=N(N-1)/2^D$.

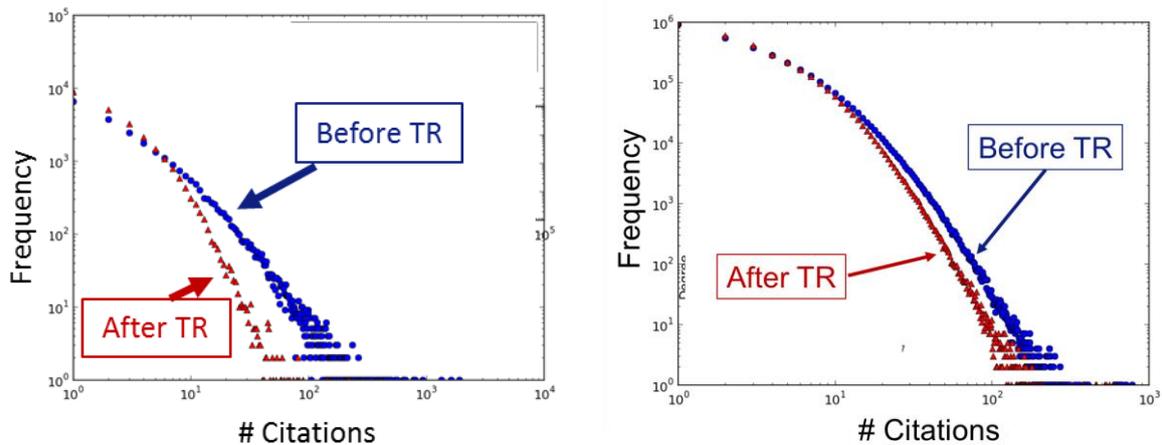

**Figure 3 The citation count distribution before and after TR. On the left the results for the quant-ph section of arXiv (paperscape dataset) shows a significant change and an overall loss of around 80% of the edges. On the other hand, US patents shown on the right lose around 15% of edge and the citation distribution remains similar.**

### Findings

One of the most striking findings is that different types of citation network show very different behaviour under TR. All the citations networks of academic papers we have studied have shown a dramatic loss in the number of edges, typically around 70% to 80%. Further, it is the high cited papers which suffer the most as can be seen in Figure 3 for the hep-th arXiv where the citation distribution becomes noticeably steeper. On investigation it is clear that the edges which remain are those with the age difference between cited and citing papers. Interestingly citations in US supreme court judgements show a similar pattern (not shown) but US patents show only a moderate loss as shown in Figure 3.

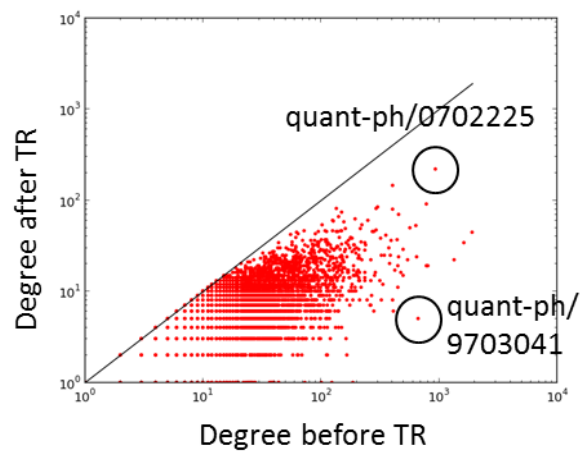

**Figure 4 The citation count before and after TR for each paper in the quant-ph paperscape data.**

Rather than looking at these bulk statistics we can look at the effect of TR on individual papers. Of course there are winners and losers. The example of the astro-ph arXiv section from paperscape.org highlights the different fates of two papers, see Figure 4. Paper quant-ph/9703041 (an older research paper on quantum entanglement) is one of the most highly cited papers with 664 citations yet TR shows that anyone using quant-ph/9703041 also took information (directly or indirectly) from five other papers. On the other hand, paper quant-ph/0702225 (a more recent review of quantum entanglement) begins with a similar number of citations, 937, yet after TR it retains 219 of these.

We have also run our dimension measures on a variety of data sets. Our results are consistent whichever of the measures we use. What emerges is that we can generally give each field a well-defined dimension and that these are significantly different. For instance Figure 5 shows how papers in two parts of the arXiv repository have distinctive dimensions. For the arXiv we have found dimensions of about for hep-th (string theory), 3 for both hep-ph (particle physics) and quant-ph (quantum physics), and around 3.5 for while astro-ph (astrophysics).

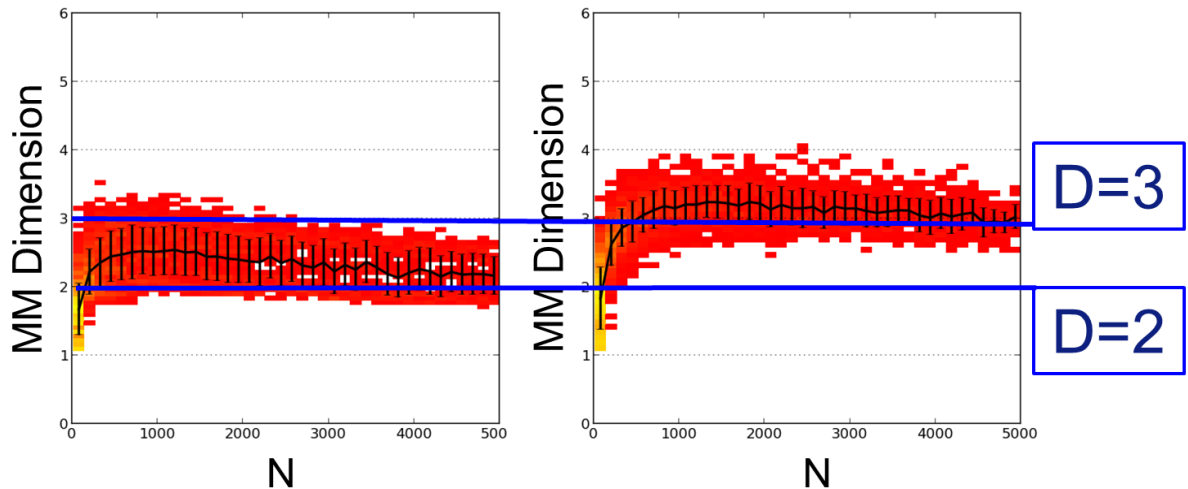

**Figure 5 Dimension of two parts of the arXiv repository (KDD cup dataset) using the MM (Myrheim-Meyer) dimension estimator. Each point represents the dimension estimated from an number of intervals defined by two randomly chosen papers. On the left the hep-th section is seen to be of lower dimension than the hep-ph section shown on the right.**

## Discussion

For us TR captures the essential causal skeleton underlying the citation network. If information is flowing from older papers to newer papers and this is reflected in the bibliographies, then all the links in the transitively reduced network are the minimum needed for such a process. Of course in practice authors may use 'short cuts' and derive information directly from older papers, but equally such short cuts were not essential and therefore there is no reason to suppose they were important. We see TR as providing a lower bound on the actual route used by the flow of important information. To go beyond this, some sort of expensive semantic analysis is needed, be it via automatic methods or by hand.

In fact we believe the transitively reduced network may be much closer to the actual set of citations of direct relevance to a publication. We have found that around 80% of links between academic papers are removed by TR. Interestingly this matches the figure given by Simkin & Roychowdhury (2003, 2005) who suggest around 80% of citations are copied from intermediate works. Any citation which was copied will always be removed by TR.

Our suggestion is that TR could be an important way to reveal which papers were essential for the developments described in a new paper. Not surprisingly, these tend to be recent papers but it is still a surprise to find such a large fraction are removed. We have shown that there are big differences in the post-TR citation count of papers in similar fields with similar high citation counts. This could be a way to discriminate between papers and could provide an alternative basis for a recommendation system. For instance searches could be ordered by post-TR citation count. One hypothesis is that papers which retain a high citation count after TR have been used across a wider range of topics. These are works which might be of more interest to researchers looking for papers outside their normal field of interest.

The behaviour of our patents and court citations also shows how TR can be a useful way to highlight different citation practices. The court data behaves in a way which is similar to that of academic papers with a large number of edges lost under TR. On the other hand, patents lose only a small fraction of their edges. The difference reflects the fact that for a patent, citations are a recognition of prior art, a legal necessity when writing a patent. However, as a patent is meant to be a novel development, they presumably try not to refer to earlier work so as to appear to be as different as possible from the literature. On the other hand, US Supreme Court judges seem to act like academic authors, citing older documents, which may have no direct relevance, along with the more recent documents, which have the latest distillation of this knowledge and are the real source of any innovation.

Our dimension measures again highlight difference between fields. We interpret the low dimension of the hep-th arXiv to suggest that string theory is a rather narrow field feeding off a few strands of research, at least when compared to hep-ph, quant-ph and astro-ph where research appears to be moving in a wider range of directions.

## Conclusions

We have argued that citation networks require a new type of measure which takes account of the constraint imposed by time. We have given some examples of how this can be done and shown that they reveal some interesting features in real citation networks. We hope to add other measures and to improve the interpretation of our results by comparing them with non-network derived measures.

## Acknowledgments

We would like to thank Damien George and Robert.Knegjens who provided us with access to their paperscape.org arXiv citation data. We also would like to acknowledge useful conversations with K.Christensen, J.Gollings, A.Hughes and T.Loach.